\documentclass{article}

\usepackage{pgfplots}
\pgfplotsset{compat=newest}
\usetikzlibrary{patterns}

\usepackage{arxiv}
\usepackage{tabularx}

\usepackage[utf8]{inputenc} 
\usepackage[T1]{fontenc}    
\usepackage{hyperref}       
\usepackage{url}            
\usepackage{booktabs}       
\usepackage{amsfonts}       
\usepackage{nicefrac}       
\usepackage{microtype}      
\usepackage{lipsum}

\usepackage{biblatex}
\title{Design of Transformation Initiatives Implementing Organisational Agility - An Empirical Study}

\author{
  Ivan Kovynyov\thanks{Corresponding author}
    \\
  Imperial College Business School\\
  South Kensington Campus \\
  London SW7 2AZ\\
  \texttt{ivan.kovynyov18@alumni.imperial.ac.uk} \\
   \And
   Axel Buerck \\
   kobaltblau Management Consultants GmbH\\
   Munich, Germany \\
   \texttt{axel.buerck@kobaltblau.com} \\
    \And
 Ralf Mikut \\
  Institute for Automation and Applied Computer Science\\
  Karlsruhe Institute of Technology\\
  Karlsruhe, Germany \\
  \texttt{ralf.mikut@kit.edu} \\
}

\begin{document}
\maketitle


\begin{abstract}
This study uses 125 responses from companies of all sizes headquartered in Germany, Switzerland, France and UK to reveal perceptions of the drivers of organisational agility. It further investigates current understanding of managing principles of multiple organisational dimensions such as culture, values, leadership, organisational structure, processes and others to achieve greater organisational agility. The data set is disaggregated into four major profiles of agile organisations: laggards, execution specialists, experimenters, and leaders. The approach to agile transformation is analysed by each of those profiles. While the positive effect from a more holistic approach is confirmed, leaders tend to focus more on processes and products rather than project work. Respondents perceive that IT, product development and research are most agile functions within their organisations, while human resources, finance and administration are considered being not agile. Further, organisations with higher levels of organisational agility tend use more than one agile scaling framework. Implications on theories of agile transformations and organisational design are discussed.
\end{abstract}


\section{INTRODUCTION}

Organisations aspire achieving greater agility, most often defined as the ability to fluidly react to changes in customer behaviour and market conditions \cite{Overby2006, Keller2019, wendler2013}. Senior executives direct attention to implementing agile practices within their organisations to improve its strategic positioning \cite{kotter2012accelerate}, improve decision making \cite{rigby2020}, and facilitate exploration of new avenues of revenue \cite{ghezzi2018agile}. Since rapid adaptation and agility improve performance in volatile environments \cite{rafique2018, drury2014}, it is important to examine what decision makers currently perceive as enabling factors of organisational agility and what are ways to achieve it.

Both agile software development and agile organisational design are theorised as influencing organisational agility, but these theoretical streams have evolved as independent literatures. In many frameworks, organisational agility arises from cultural changes, strategic flexibility and managerial practices \cite{wendler2013, Kalenda2018}. Further studies find that agile architectures \cite{leffingwell2008principles}, ways of working  \cite{Lindsjorn2016} and employee empowerment \cite{menon2001} improve the organisation's ability to respond rapidly and effectively. Studies on organisational design however attribute agility to structures that facilitate flexibility and impose changes in managerial control systems \cite{bernstein2016, kotter2012accelerate}.

Gaps nevertheless exist in understanding how companies attain organisational agility. First, it is unclear whether organisational agility comes from technical excellence or agile organisational design. Second, models examine organisational agility independent of organisation-wide transformation efforts \cite{mathiassen2006, ambrose2004}. Third, although studies have generated interesting results on designing individual dimensions of organisational agility such as structure, leadership style or software development, there is a paucity of global, multi-dimensional studies addressing interdependence across individual dimensions. Also, it lacks an empirically backed overview of best practices for design and implementation of agile transformation initiatives.

This study uses 125 responses from companies of all sizes headquartered in Germany, Switzerland, France and UK to reveal perceptions of the drivers of organisational agility. It further investigates current understanding of managing principles of multiple organisational dimensions such as culture, values, leadership, organisational structure, processes and others to achieve greater organisational agility. Our study is the first empirical effort to address three questions about designing and implementing agile transformation initiatives: What are major profiles for agile organisations? How do organisations balance efforts across individual organisational dimensions, such as structure, leadership style, culture, software development practices and project work? What are best practices in designing agile transformation initiatives?

The data set is disaggregated into four major profiles of agile organisations: laggards, execution specialists, experimenters, and leaders. The approach to agile transformation is analysed by each of those profiles. While the positive effect from a more holistic approach is confirmed, leaders tend to focus more on processes and products rather then project work. Respondents perceive that IT, product development and research are most agile functions within their organisations, while human resources, finance and administration are considered being not agile. Further, organisations with higher levels of organisational agility tend use more than one agile scaling framework. These findings inform an important area of managerial practices and present opportunities for future research.



\section{THEORY}
\label{sec:theory}

\subsection{Literature Review and Research Questions}
\label{subsec:lit-review}

\textit{Technical Excellence vs. Agile Organisational Design
}

Agility involves organisation's responsiveness to changes \cite{Overby2006} and a proactive rather than reactive attitude. Early studies of agility relied on observations from self-governed, autonomous software engineering teams \cite{reich1999building} and from process improvements within manufacturing systems \cite{vokurka1998journey, sharifi1999methodology, takeuchi1986}. However, more recent studies refer to agility not only as an outcome of technological achievement but rather as a result of human ability, skills and motivation \cite{sherehiy2007}. Shifting this understanding from rather a technological implementation to an enterprise management system has reframed agility as an organisational agility.

A well-examined characteristic associated with organisational agility is agile organisational design. Novel organisational forms facilitating value orientation and cross-functional work arise: value streams \cite{rother2003}, Holacracy \cite{Robertson2015, bernstein2016}, DevOps \cite{ebert2016} and others. Similarly, new structures require new roles and responsibilities: agile coach \cite{davies2009} and product owner \cite{bass2015}. Agile scaling frameworks lay foundations for implementing agility on an organisation-wide level: SAFe \cite{leffingwell2018safe}, LeSS \cite{larman2016large}, Spotify model \cite{Kniberg2012}, Scrum of Scrums \cite{Sutherland2005} and others.

From a technical perspective, organisational agility is determined by agile software development practices \cite{Beck2001, Martin2002}. Studies suggest that frequent delivery, small batch size, agile requirements engineering and agile testing procedures are essential technical antecedents for organisational agility \cite{Chow2008, deSouza2014}. Further, team diversity and autonomy are crucial for success in agile teams \cite{Lee2010, Lindsjorn2016}.

However, it is unclear whether organisational agility comes from agile organisational design alone or it requires appropriate technical excellence. Also, there is a paucity of in-depth analyses of mechanisms imposing organisational agility through technical improvements and organisational changes.

Recent studies suggest that organisational agility arises from corporate values, technology, change management practices, agile collaboration styles and structures \cite{wendler2012}. It appears to have a more subtle relationship to individual organisational dimensions; therefore we investigate how organisations balance efforts across those dimensions to thrive organisational agility.

Finally, some scholars argue that agility supports generating value from digital technologies \cite{kovynyov2019digital, ghezzi2020agile}. Therefore we briefly elaborate on this argument by investigating the relationship between organisational agility and digital initiatives.

\textit{Scaling Agile}


Previous studies suggest that many benefits of organisational agility derive from scaling agile practices \cite{Kalenda2018}. Therefore the demand for agile scaling frameworks has increased \cite{Rigby2016}. Some papers investigate scaling agile methods in large software development projects \cite{Alqudah2016}, other focus on using agile methods in large-scale product development initiatives \cite{Kettunen2008}. Increase in the agile scaling effort creates need for the selection criteria \cite{diebold2018scaling}.

In line with prior research, we refer to agile scaling frameworks as conceptual frameworks implementing agile values, principles and practices on the enterprise-wide level, for instance, scaled agile framework SAFe \cite{leffingwell2018safe}, Spotify model \cite{Kniberg2012}, Scrum of Scrums \cite{Sutherland2005} and others. We expect a positive relationship between using agile scaling frameworks and the level of organisational agility and address the current usage of frameworks in corporate environments.

\textit{Agile Transformation Initiatives and Measuring Agility}

Research has assessed challenges and success factors of agile transformation initiatives \cite{dikert2016challenges}. Although agile transformation design has been studied in context of large-scale software development projects \cite{Alqudah2016} and product development initiatives \cite{Kettunen2008}, there is less empirical research in context of companies seeking agility on the organisation-wide level. Recent studies suggest complex relationships between individual organisational dimensions when implementing and measuring agile transformation effort \cite{Kettunen2008}. Several maturity models are used to understand and measure those relationships \cite{wendler2013, gren2015}.

In line with prior research, we expect that individual organisational dimensions are affected by agile transformations in different ways and investigate this relationship by identifying major profiles of agile organisations. Finally, we consider the design of agile transformation initiatives and derive best practices when designing and implementing such initiatives.

\subsection{Suggested Framework}
\label{subsec:framework}




Research confirms impact of agile practices on multiple organisational dimensions \cite{sherehiy2007, wendler2013}. Some scholars suggest that organisational agility impacts corporate values, technology, change management practices, collaboration styles and organisational structure \cite{wendler2012}. Other studies draw relationship to architecture \cite{leffingwell2008principles}, ways of working  \cite{Lindsjorn2016} and people management \cite{menon2001}. Further, studies on organisational design attribute agility to structures that facilitate flexibility and impose changes in managerial control systems \cite{bernstein2016, kotter2012accelerate}. We suggest therefore the following six organisational dimensions to assess the impact of organisational agility on organisations:
 (i) culture, values and leadership, (ii) organisation and structure, (iii) delivery and software development, (iv) product development, (v) ways of working, and (vi) enterprise architecture.

\textit{Culture, values and leadership} cover leadership and operating styles of the management, norms and behaviours people follow across the organisation, how people interact at work with each other within the organisation and with external partners such as clients and vendors \cite{bradach1996}. Related agile practices and tools include agile goal setting using Objectives \& Key Results method (OKR) \cite{niven2016}, agile leadership practices \cite{baker2007}, continuous improvement with Kaizen \cite{berger1997}, feedback culture \cite{strode2009}, employee empowerment \cite{menon2001}, self-organisation, Management 3.0 practices \cite{appelo2016}, agile mindset, fail-faster-principle, agile coaching \cite{davies2009} and others. Key differentiator between agile and non-agile organisations for this domain is the attitude towards risk-taking. Agile organisations consider failure is an essential part of learning and embrace taking calculated risks, while traditional organisations usually follow plan-and-execute approaches cultivating zero-failure-tolerance \cite{strode2009}.

\textit{Organisation and structure} refer to ways "in which tasks and people are specialised and divided, and authority is distributed" across the organisation \cite{bradach1996}. This dimension includes grouping of activities and reporting relationships into organisational units, formal and informal procedures and processes used to manage the organisation. Related agile practices and tools are  cross-functional teams \cite{parker2003}, new agile roles (agile coaches \cite{davies2009}, product owners \cite{bass2015} and others), novel organisational forms (value streams \cite{rother2003}, Holacracy \cite{Robertson2015, bernstein2016}, DevOps \cite{ebert2016} and others), and agile scaling frameworks (SAFe \cite{leffingwell2018safe}, LeSS \cite{larman2016large}, Spotify model \cite{Kniberg2012} and others). With that in mind, the level of cross-functional collaboration can be seen as the major differentiator between agile and non-agile organisations. In agile organisations, individual functional parts collaborate seamlessly across divisions of an organisation to create value. General work is organised in a cross-functional manner rather than in functional silos. The collaboration mode reconfigures fluidly and adapts to the changing environment. Traditional Tayloristic organisations however exhibit behaviours where individual functional parts follow their own agendas and focus on local optima rather than improving the entire system.

\textit{Delivery and software development} include all activities associated with implementing new software solutions within the organisation such as software development life cycle, project management approach and maintenance procedures. This domain focuses in particular on ways of organising large-scale software projects such as introduction of a application, relaunch or replacement of existing systems and applications. Here, organisations tend to use agile practices and tools such as extreme programming \cite{Beck2001}, disciplined agile delivery \cite{ambler2012}, test-driven development \cite{beck2003testdriven}, test automation \cite{collins2012software}, continuous delivery \cite{humble2010continuous}, pair programming \cite{vanhanen2007experiences}, minimal viable products \cite{lenarduzzi2016mvp}, minimal marketable features \cite{cleland2005mmf} and others. Considering the project work, the amount of up-front planning can be considered as key differentiator. While traditional organisations spend significant amount of work for high-level planning activities before kicking-off the project, agile organisations distribute those activities over the entire project duration. The batch size (amount and magnitude of software changes in one release cycle) and release frequency can be seen as key differentiators between agile and non-agile organisations. Agile organisations tend to use processes and procedures allowing them to release small pieces of software frequently. Traditional organisations usually follow a more fixed schedule of major monthly or quarterly releases.

\textit{Product development} refers to creation and launch of new products that satisfy a newly created customer need or market niche. This dimension also covers modification of existing products. Related agile practices include customer journey mapping, design thinking \cite{liedtka2018design}, customer centricity \cite{shah2006path}, design sprints \cite{banfield2015design}, Lean Start-up methodology \cite{reis2011lean} and others. Distance to the customer is key differentiator \cite{Rigby2016}. Agile organisations are able to early capture changes in customer needs and fluidly reconfigure.

\textit{Ways of working} cover practices of organising, performing, leading, along with new approaches to recruiting, developing and engaging employees. Ways of working usually refer to Scrum \cite{schwaber2002agile}, Kanban \cite{ahmad2013kanban}, Kaizen \cite{berger1997}, agile retrospectives \cite{derby2006agile}, Beyond Budgeting \cite{libby2010beyond} and others. Levels of employee autonomy and amount of decision rights a regular employee is equipped with can be considered as differentiating factors in this domain.

\textit{Enterprise architecture} refers to the fundamental orchestration of software systems and its components, their relationship to each other and towards the external environment as well as general principles of governance relating to design and evolution of those systems \cite{winter2006}. This organisational dimension further includes a set of values, principles and practices that support active, evolutionary design and architecture of the organisation's systems landscape. Here, organisations use such practices as architectural runway \cite{buchmann2012architectural}, agile architecture \cite{leffingwell2008principles} and others. Agile architectures are more federated \cite{leffingwell2008principles}. Organisations are seeking to solve a trade-off between organisational agility and reliability through decoupling \cite{Keller2019}. They have loosely coupled, independent modules, while traditional architectures mostly contain of monolithic legacy systems.

We formalise the differences between agile and non-agile organisations by introducing maturity levels for each dimension. The score ranges from 1 to 5 with 1 being the lowest and 5 being the highest scores. The maturity levels are defined as follows: (1) non-agile, (2) selected basic agile principles and tools are implemented, (3) core agile principles and tools are implemented, (4) advanced agile principles and tools are implemented, and (5) front-running, novel agile tools and practices are piloted.



\section{DATA AND METHOD}
\label{sec:data-collection}

\subsection {Target group and distribution}

The target group consists of senior executives, business leaders and agile practitioners in small, medium-size and large enterprises predominantly in Europe, regardless of industry or particular business area. Also, the target group includes organisations which have recently undertaken an agile transformation. The respondents should have gained experience in applying agile principles and practices across their organisations in the recent past, for instance, through participation in agile transformation programmes as a sponsor, agile coach, change manager, line manager or senior executive. Respondents can be part of internal IT organisation, but also work for the business units. We worked with kobaltblau Management Consultants to compile the contact list that fits the selected profile. Candidates have been randomly selected from the provided list. The survey took place in April - May 2019.


It was a voluntary survey and was conducted by means of an digital questionnaire. The questionnaire included ten core and six optional questions (for classification purposes) and can be found in Section \ref{subsec:survey-questions}. We created questions in three languages: English, German, and French. Participants have been given the possibility to switch across the languages at any point in time, after they have started the survey. A number of measures has been implemented to prevent survey taking fatigue: (i) one question at a time appeared on screen, (ii) questions numbers were hided, and (iii) the overall progress bar was set up to indicate percentage completed.


The questionnaire was delivered to the target group by (i) approaching the audience individually with a personalised email including the direct link to the survey (invite over email) and (ii) by sending out standardised mass emails to the target group (mass mailing).

Table \ref{tab:response-rates} reports a breakdown of the data sample response rate by distribution type. We approached the audience of 1,044~persons and obtained 210 responses. 85 out of 210 responses appeared to be partial responses (participants have not finished answering the questionnaire) and were excluded from further consideration. The final sample data set included therefore 125 responses.
Higher response rates were obtained from individual invites over email.
The response rate of individual invites was 33 per cent, while only 3 per cent of mass mail receivers responded to the invitation.
Same for the completion rate: 61 per cent of the participants that have been approached individually and responded to the invitation have completed the survey, while the completion rate for the mass mail receivers was only 40 per cent.
  \begin{table}[hbt]
  \small
 \caption{Key response statistics by distribution type}
  \centering
  \begin{tabular}{lrrr}
    & Mass mailing & Invite over email & Total \\
    \midrule
   Audience size & 444 & 600* & 1,044 \\
   Total responses, thereof & 13 & 197 & 210 \\
   { } partial responses & 8 & 77 & 85 \\
   { } finished responses & 5 & 120 & 125 \\
   \midrule
   Response rate & 3\% & 33\% \\
   Completion rate & 40\% & 61\% \\
    \bottomrule
    *estimate
  \end{tabular}
  \label{tab:response-rates}
 \end{table}





\subsection{Respondent Profile}

Table \ref{tab:respondents-profile} reports a summary profile of respondents including (i) the size of organisation measured in terms of the total number of current employees, (ii) geography measured by the country of headquarter, (iii) industry sector of the organisation, and (iv) the job function of the responding person. The data sample includes predominantly large enterprises, as almost 40 per cent of responses represent organisations with the total of employees exceeding 10,000. About 85 per cent of the respondents work for organisations headquartered across Western Europe, in Switzerland and UK. Germany is the major country included in the sample covering almost 50 per cent of responses. Major industry sectors are the financial services (21\%), transport and logistics (14\%), high tech (10\%), and automotive (9\%). We observed a high rate of senior management participation (60 per cent) in the survey covering such positions as chief information officer (CIO), chief financial officer (CFO), chief executive officer (CEO), chief digital officer (CDO), board member, executive director, director, and business unit head. Other 40 per cent cover (senior) expert positions and roles.
 \begin{table}[hbt]
   \small
 \caption{Summary profile of respondents}
  \centering
  \begin{tabular}{lll}
  Breakdown & Responses & Percentage \\
    \midrule
   Sample size & 125 & 100\% \\
   \textbf{Size in employees} \\
    { } Fewer than 500 & 33 & 26\% \\
    { } 500-999 & 9 & 7\% \\
    { } 1,000-4,999 & 21 & 17\% \\
    { } 5,000-9,999 & 13 & 10\% \\
    { } 10,000 or more & 49 & 39 \% \\
    \midrule
    \textbf{Countries of headquarter} \\
    { } Germany & 63 & 50\% \\
    { } Switzerland & 23 & 18\% \\
    { } France & 15 & 12\% \\
    { } Belgium & 3 & 2\% \\
    { } USA & 3 & 2\% \\
    { } UK & 2 & 2\% \\
    { } Other & 16 & 13\% \\
    \midrule
    \textbf{Key industry sectors}\\
    { } Financial services (banking, insurance, and asset management) & 26 & 21\% \\
    { } Transport and logistics & 18 & 14\% \\
    { } High tech & 12 & 10\% \\
    { } Automotive & 11 & 9\% \\
    { } Manufacturing & 8 & 6\% \\
    { } Consumer goods & 7 & 6\% \\
    { } Communication, media and entertainment & 7 & 6\% \\
    { } Energy and utilities & 5 & 4\% \\
    { } Public sector & 4 & 3\% \\
    { } Miscellaneous (healthcare, pharmaceuticals, chemicals and others) & 27 & 22\%  \\
    \midrule
   \textbf{Job function} \\
    { } IT manager & 72 & 58\% \\
    { } Business manager & 46 & 37\% \\
    { } Agile coach & 7 & 6\% \\
    \bottomrule
    Note: percentages may not add up to 100 due to rounding
  \end{tabular}
  \label{tab:respondents-profile}
 \end{table}

\subsection{Data Analysis}


Our data analysis approach has been primarily designed to examine correlations among the scores for six organisational dimensions introduced in the Section \ref{subsec:framework}. In order to create comprehensive visuals and simplify interpretation of our results, we mapped the sample to a low-dimensional representation using a PCA (Principal Component Analysis) method. We found no evidence for violation of normal distribution assumptions in the data set, therefore we chose PCA as a simple and efficient method for dimensional reduction to generate aggregated features. We used normalised dimensional scores with mean $\mu = 0$ and standard deviation $\sigma = 1$.

The samples of the six organisational dimensions were clustered using the Fuzzy C-Means method \cite{Bezdek81} with the fuzzifier-value of 1.1 and a selection of the number of clusters with the separation measure. Fuzzy C-means was chosen as a robust clustering method with a stable convergence behaviour towards similar solutions. Also, Fuzzy C-means can process gradual memberships of participants to the different clusters during the cluster generation. For the sake of simplicity, participants were assigned to the cluster by the highest membership score.



We used Qualtrics for the data collection and initial data processing purposes. For the PCA analysis and clustering process, we used the MATLAB toolbox SciXMiner \cite{Mikut17}. Selected visuals and the interactive data room were developed with  Tableau.

\section{RESULTS}
\label{sec:results}

\subsection{Four Profiles of Agile Organisations}
\label{subsec:four-profiles}

Analysis of the dimensional scores reveals moderate positive correlations across all dimensions (see Table~\ref{tab:corr}). The lowest correlations values were found between D2 (Organisation and structure) to D3 (Delivery and software development) and D2 to D4 (Product development) supporting the idea that the organisations either choose improving the technical space with agility or addressing organisational changes.

We mapped six dimensional scores into a two-dimensional feature space using the aggregated features. The first component is defined as the weighted mean of all dimensional scores and labeled as 'Aggregated Feature 1' (AF1), exploiting the positive correlations across all dimensions. The second component focuses on the difference between the dimensional scores for D1 and D2 against other dimensional scores. We labeled this component as 'Aggregated Feature 2' (AF2).
These differences are highlighted by the positive and negative signs of the correlations between the second component and D1 and D2 vs. D3-D6 (see Table~\ref{tab:corr}). First and second PCA components explain 49 respectively 16.5 per cent of the total variance.
\begin{table}[hbt]
   \small
 \caption{Pearson correlation coefficients for the dimensional scores and aggregated features}
  \centering
  \begin{tabular}{lcccccc}
  Feature & D1 & D2 &D3 &D4 &D5 &D6  \\
   \midrule
    D1: Culture, values and leadership & 1.00& & & & & \\
    D2: Organisation and structure & 0.49& 1.00&  & & &  \\
    D3: Delivery and software development  & 0.40 & 0.19 & 1.00& & &  \\
    D4: Product development & 0.41& 0.12 & 0.43& 1.00& & \\
    D5: Ways of working  & 0.41& 0.41& 0.52& 0.42& 1.00& \\
    D6: Enterprise architecture & 0.29& 0.26& 0.43& 0.38& 0.56& 1.00\\
    \midrule
    PCA1: Aggregated Feature 1  & 0.71& 0.56& 0.72& 0.66& 0.81& 0.71 \\
    PCA2: Aggregated Feature 2  & 0.36& 0.75& -0.31& -0.38& -0.04& -0.24\\
\end{tabular}
  \label{tab:corr}
 \end{table}



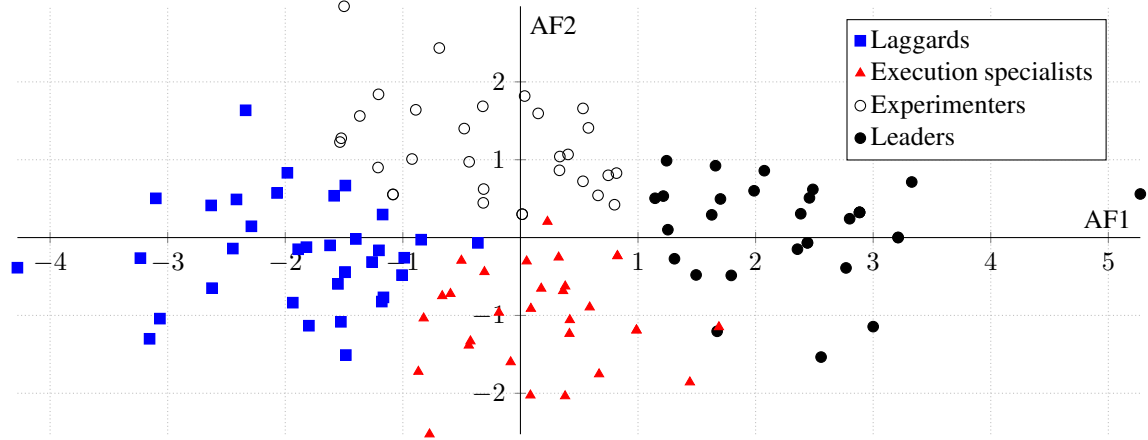
\begin{figure}
\centering
\caption{Scatter plot with Aggregated Features 1 and 2}
\label{fig:scatter-plot}

\begin{tikzpicture}
\begin{axis}[
scatter/classes={
    a={mark=square*,blue}, 
    b={mark=triangle*,red}, 
    c={mark=o,draw=black}, 
    d={mark=*,black}}, 
legend style={ cells={anchor=west}, legend pos=north east},
enlargelimits=0.3,
axis lines=middle,
axis line style={-},
xlabel=AF1,
ylabel=AF2,
grid=major,
grid style={densely dotted},
width=\textwidth,
height=\axisdefaultheight
]


\addplot[scatter,only marks,
scatter src=explicit symbolic] coordinates {

(-0.844,	-0.027)	[a]
(-1.165,	-0.768)	[a]
(-1.172,	0.296)	[a]
(-1.007,	-0.482)	[a]
(-1.492,	-0.445)	[a]
(-1.492,	-0.445)	[a]
(-1.89,	    -0.15)	[a]
(-0.362,	-0.068)	[a]
(-0.988,	-0.259)	[a]
(-1.179,	-0.822)	[a]
(-1.262,	-0.314)	[a]
(-1.203,	-0.164)	[a]
(-1.4,	    -0.018)	[a]
(-1.619,	-0.1)	[a]
(-1.552,	-0.595)	[a]
(-1.485,	-1.509)	[a]
(-1.489,	0.668)	[a]
(-1.527,	-1.082)	[a]
(-1.584,	0.537)	[a]
(-1.799,	-1.132)	[a]
(-1.936,	-0.836)	[a]
(-2.07,	    0.573)	[a]
(-1.82,	    -0.123)	[a]
(-1.982,	0.832)	[a]
(-2.415,	0.49)	[a]
(-2.337,	1.635)	[a]
(-2.446,	-0.141)	[a]
(-2.288,	0.145)	[a]
(-3.232,	-0.264)	[a]
(-3.101,	0.504)	[a]
(-2.63,	    0.413)	[a]
(-3.154,	-1.3)	[a]
(-3.066,	-1.041)	[a]
(-2.623,	-0.65)	[a]
(-4.278,	-0.387)	[a]

(5.272,	0.56)	[d]
(3.212,	0.001)  [d]
(3.212,	0.001)	[d]
(3.328,	0.714)	[d]
(2.884,	0.324)	[d]
(2.884,	0.324)	[d]
(2.884,	0.324)	[d]
(2.769,	-0.39)	[d]
(2.8,	0.242)	[d]
(2.441,	-0.067)	[d]
(2.441,	-0.067)	[d]
(2.487,	0.619)	[d]
(2.384,	0.306)	[d]
(2.356,	-0.149)	[d]
(1.659,	0.923)	[d]
(1.701,	0.496)	[d]
(1.627,	0.292)	[d]
(2.074,	0.859)	[d]
(1.793,	-0.486)	[d]
(1.987,	0.601)	[d]
(1.215,	0.533)	[d]
(1.145,	0.505)	[d]
(1.31,	-0.272)	[d]
(3,	    -1.144)	[d]
(2.557,	-1.535)	[d]
(2.458,	0.51)	[d]
(1.494,	-0.48)	[d]
(1.673,	-1.204)	[d]
(1.243,	0.987)	[d]
(1.254,	0.101)	[d]

(0.825,	-0.236)	[b]
(0.825,	-0.236)	[b]
(1.688,	-1.149)	[b]
(1.441,	-1.858)	[b]
(0.987,	-1.19)	[b]
(0.987,	-1.19)	[b]
(0.417,	-1.235)	[b]
(0.325,	-0.254)	[b]
(0.23,	0.205)	[b]
(0.589,	-0.895)	[b]
(0.42,	-1.058)	[b]
(0.381,	-0.626)	[b]
(0.089,	-0.913)	[b]
(-0.083,	-1.599)	[b]
(0.177,	-0.654)	[b]
(0.053,	-0.304)	[b]
(-0.823,	-1.036)	[b]
(-0.502,	-0.295)	[b]
(-0.595,	-0.722)	[b]
(-0.665,	-0.75)	[b]
(-0.439,	-1.385)	[b]
(-0.306,	-0.441)	[b]
(-0.183,	-0.963)	[b]
(-0.425,	-1.331)	[b]
(0.363,	-0.685)	[b]
(0.669,	-1.754)	[b]
(0.38,	-2.036)	[b]
(0.085,	-2.026)	[b]
(-0.773,	-2.526)	[b]
(-0.868,	-1.722)	[b]


(0.817,	0.828)	[c]
(0.66,	0.542)	[c]
(0.747,	0.8)	[c]
(0.332,	0.864)	[c]
(0.405,	1.068)	[c]
(0.335,	1.041)	[c]
(0.578,	1.409)	[c]
(0.532,	0.723)	[c]
(0.015,	0.3)	[c]
(0.015,	0.3)	[c]
(0.036,	1.818)	[c]
(-0.436,	0.973)	[c]
(-0.478,	1.4)	[c]
(-0.316,	0.446)	[c]
(-0.32,	1.686)	    [c]
(-0.313,	0.623)	[c]
(-0.922,	1.009)	[c]
(-1.084,	0.554)	[c]
(-1.084,	0.554)	[c]
(0.8,	0.422)	[c]
(0.532,	1.659)	[c]
(0.151,	1.595)	[c]
(-0.69,	2.435)	[c]
(-0.89,	1.641)	[c]
(-1.211,	0.9)	[c]
(-1.207,	1.84)	[c]
(-1.5,	2.971)	[c]
(-1.535,	1.227)	[c]
(-1.366,	1.562)	[c]
(-1.524,	1.276)	[c]

};

\legend{Laggards, Execution specialists, Experimenters, Leaders}
\end{axis}
\end{tikzpicture}

\end{figure}

Figure~\ref{fig:scatter-plot} shows the values of the Aggregated Features 1 and 2 as a scatter plot. The dots represent values of Aggregated Feature 1 (x-axis) and Aggregated Feature 2 (y-axis). The position of each dot on the horizontal and vertical axes indicate each individual response in the survey.
The visual analysis of the scatter plot leads to impression that the clusters have blurred boundaries. However, the visual representation appeared to be useful in discussing the positioning of individual responses relative to its peers by industry, geography or company size.

Cluster analysis based on six organisational dimensions resulted in four clusters which we interpreted as four empirical profiles of agile organisations. Table \ref{tab:clusters} describes these four clusters by using cluster mean values of six dimensional scores. Based on the interpretation of the dimensional scores, we labeled the clusters as Laggards (lowest scores across all dimensions), Execution Specialists (high scores for delivery and software development, other scores around sample average), Experimenters (high scores for organisation and structure, above average score for culture, values and leadership, other scores around sample average), and Leaders (highest scores across all dimensions).
 \begin{table}[hbt]
  \small
 \caption{Breakdown of dimensional scores by cluster}
  \centering
  \begin{tabular}{lrrrr}
     & Laggards & Experimenters & Execution specialists & Leaders \\
    \midrule
    Cluster size & 35 & 30 & 30 & 30 \\
    \midrule
    D1: Culture, values and leadership & 2.1 & 2.9 & 2.4 & 3.8 \\
    D2: Organisation and structure & 2.0 & 3.7 & 2.2 & 3.7 \\
    D3: Delivery and software development & 2.3 & 2.2 & 3.1 & 4.1 \\
    D4: Product development & 2.4 & 2.5 & 3.7 & 4.1 \\
    D5: Ways of working & 1.6 & 2.6 & 2.7 & 3.6 \\
    D6: Enterprise architecture & 1.7 & 2.2 & 3.1 & 3.4 \\
    \midrule
    Overall level of organisational agility & 2.0 & 2.7 & 2.9 & 3.8 \\
  \end{tabular}
  \label{tab:clusters}
 \end{table}

\textbf{Laggards} (n=35): Agile practices and DevOps elements are piloted across organisation as isolated spots, especially within the IT and R\&D departments. No enterprise-wide agile culture established. Overall organisational structure remains unchanged. Project-orientated thinking prevails. No product orientation.

\textbf{Experimenters} (n=30): High scores across the dimensions relating to corporate culture, values, leadership and organisation. Novel organisational forms, cross-functional collaboration and agile processes are on top of the agenda. Clear focus on people rather than technical maturity. Organisations within this clusters are more likely to start agile transformations on the business side or from the people's perspective.

\textbf{Execution specialists} (n=30): Agile tools and practices are well established and support the agile delivery model. Clear product orientation and working along the value streams are institutionalised. Structure, roles and responsibilities remain Tayloristic. The tipping point in terms of culture, values and organisation is not reached.

\textbf{Leaders} (n=30): Consistently high scores across all dimensions. High level of customer integration based on agile delivery model. Tipping point across culture, values and organisation clearly reached. Agile ways of working are dominating across the organisation. Agile delivery model includes product orientation, short time-to-market and frequent customer feedback cycles.










\subsection{Design of Agile Transformation Initiatives}

\subsubsection{Affected Organisational Dimensions}

The survey asked participants to point out key organisational dimensions affected most by the transformation efforts within in their organisations (see Table \ref{tab:most-affected-dimensions}).
Respondents perceive that project management, delivery and software development, processes, and product development are impacted most by agile transformation initiatives. However, while the laggards, experiments and execution specialists agree on project management being impacted most, the leaders direct attention to processes and product management instead. Also, leaders have higher number of responses relating to culture and values as well as the goal setting approach.

Table \ref{tab:most-affected-dimensions} further reports average numbers of selected dimensions by response. On average,  participants have selected 3.7~dimensions. While the experimenters and execution specialists hover around the sample average, laggards have a lower value of 3.3. The leaders stand out with 4.4 supporting the idea that this cluster seeks a more holistic approach with a greater organisational reach compared to another clusters. In another words, companies with above-average levels of organisational agility tend to design their agile transformation initiatives with a greater organisational reach by tacking a larger number of organisational dimensions.
%
\begin{table}[hbt]
 \small
 \caption{Which dimensions of your organisation have been affected most by agile transformation [count responses]}
  \centering
  \begin{tabularx}{\textwidth}
  {lXXXXX}
    Dimension & Laggards & Experimenters & Execution specialists & Leaders & Total \\
    \midrule
    Project management & 18 & 18 & 20 & 20 & 76 \\
    Delivery and software development & 16 & 16 & 15 & 16 & 63 \\
    Processes & 15 & 14 & 13 & 20 & 62 \\
    Product development & 13 & 9 & 10 & 15 & 62 \\
    Leadership & 11 & 10 & 9 & 11 & 41 \\
    Culture and values & 7 & 10 & 7 & 13 & 37 \\
    Organisational structure & 4 & 13 & 9 & 9 & 35 \\
    Software maintenance & 9 & 4 & 5 & 7 & 25 \\
    Goal setting & 3 & 4 & 7 & 10 & 24 \\
    Demand management & 4 & 6 & 4 & 4 & 18 \\
    Architecture & 5 & 3 & 4 & 3 & 15 \\
    Governance & 5 & 2 & 1 & 3 & 11 \\
    Portfolio Management & 4 & 2 & 4 & 1 & 11 \\
    \midrule
    Average number of dimensions per response & 3.26 & 3.70 & 3.60 & 4.40 & 3.72 \\
  \end{tabularx}
  \label{tab:most-affected-dimensions}
 \end{table}

\subsubsection{Reported Share of Agile Projects in the Project Portfolio}


Table~\ref{tab:agile-projects} summarises reported shares of agile projects in the project portfolios of the respondents.
Our results confirm that companies enacting higher levels of organisational agility have higher share of agile projects in their portfolios. Indeed, 60 per cent of the leaders report the share of agile projects of 60 per cent or more. However, our data do not allow us to confirm the underlying causal mechanisms between organisational agility and share of agile projects. Indeed, high share of agile projects might lead to higher levels of organisational agility. Or vice versa, high organisational agility might be a cause to higher shares of agile projects. Further research can more clearly delineate the underlying reasons of this relationship.
\begin{table}[hbt]
 \small
 \caption{Reported share of agile projects in the project portfolio by cluster}
  \centering
  \begin{tabular}{llllll}
     & <20\% & 20-40\% & 40-60\% & >60\% & Total \\
    \midrule
    Laggards & 19 & 11 & - & 5 & 35 \\
    Experimenters & 6 & 10 & 7 & 7 & 30 \\
    Execution specialists & 13 & 10 & 2 & 5 & 30 \\
    Leaders & 3 & 6 & 3 & 18 & 30 \\
    \midrule
  \end{tabular}
  \label{tab:agile-projects}
 \end{table}

\subsubsection{Reported Length of Experience with Agile Methods}


Table \ref{tab:length-experience} summarises reported length of experience with agile methods by cluster. The vast majority of participants confirmed having experience with agile methods of less than four years: laggards (71\%), experimenters (83\%), execution specialists (76\%), and leaders (50\%). Leaders seem to have longer experience with agile methods compared to another clusters supporting the idea that greater organisational agility comes along with experience. However, our data do not allow us to conclude that longer experience does lead to greater organisational agility or vice versa.
%
%
\begin{table}[hbt]
 \small
 \caption{Length of experience with agile practices by cluster}
  \centering
  \begin{tabular}{llllll}
     & <2 years & 2-4 years & 4-6 years & >6 years & Total \\
    \midrule
    Laggards & 16 & 9 & 5 & 5 & 35 \\
    Experimenters & 11 & 14 & 3 & 2 & 30 \\
    Execution specialists & 14 & 9 & 3 & 4 & 30 \\
    Leaders & 5 & 10 & 6 & 9 & 30 \\
     \midrule
  \end{tabular}
  \label{tab:length-experience}
 \end{table}

\subsubsection{Perceived Level of Adoption of Agile Methods Relative to Competition}

The survey asked participants to report the level of adoption of agile methods relative to competitors (see Table \ref{tab:adoption-level-competitors}). The perception of respondents varies significantly by cluster. The laggards predominantly consider being at the same level or worse compared to the competition (86\%). Only 14 per cent of the laggards rate themselves better than the competition in terms of using agile methods. 40 per cent of the execution specialists and 43 per cent of the experimenters report having a higher level of adoption compared to the competition. Finally, 70 per cent of the leaders feel having a higher level of adoption. The general perception seems to be realistic: organisations exhibiting lower levels of organisational agility tend to grade themselves below the competition and vice versa. The respondents with higher levels of organisational agility report more confidence and satisfaction from using agile methods. We haven't found any statistically significant deviation by industry, geography or company size.
\begin{table}[hbt]
 \small
 \caption{Perceived level of adoption of agile practices relative to competitors [count responses]}
  \centering
  \begin{tabular}{lllllll}
     & Much better & Somewhat better & About the same & Somewhat worse & Much worse & Total \\
    \midrule
    Laggards & - & 5 & 14 & 10 & 6 & 35 \\
    Experimenters & 3 & 10 & 13 & 3 & 1 & 30 \\
    Execution specialists & 3 & 9 & 9 & 7 & 2 & 30 \\
    Leaders & 9 & 12 & 7 & 2 & - & 30 \\
     \midrule
  \end{tabular}
  \label{tab:adoption-level-competitors}
 \end{table}

\subsubsection{Relation to Digitalisation Initiatives}

Table \ref{tab:digital-initiatives} summarises the usage of digitalisation initiatives by organisations from our data set. Multiple answers were allowed. Digitalisation projects and initiatives are at the top of the agenda for all four groups, followed by digital strategy, except for the laggards focusing on automating business processes instead. Further, the vast majority of respondents direct attention to developing digital products and services; however, the leaders seem to connect those initiatives with developing novel digital business models.

Further, following current discussion on the role of a chief digital officer \cite{wladawsky2012, tumbas2018, haffke2016}, our data confirm low level of adoption for this role. Clusters with higher level of organisational agility seem to avoid this role supporting the idea that organisations are more effective when digital competence is incorporated into the DNA of the entire organisation rather than concentrated in one particular unit.
\begin{table}[hbt]
 \small
 \caption{Current use of digital initiatives [count responses]}
  \centering
  \begin{tabularx}{\textwidth}
  {lXXXXX}
     & Laggards & Experimenters & Execution specialists & Leaders & Total \\
    \midrule
    Digital transformation projects and initiatives & 27 & 23 & 22 & 23 & 95 \\
    Digital strategy & 15 & 23 & 19 & 20 & 77 \\
    Digital products and services & 19 & 15 & 14 & 19 & 67 \\
    Automated business processes & 16 & 14 & 14 & 16 & 60 \\
    Digital business models & 11 & 14 & 8 & 17 & 50 \\
    Chief digital officer & 12 & 10 & 7 & 5 & 34 \\
    Digital factory & 8 & 9 & 8 & 3 & 28 \\
    None from above & 3 & 2 & 2 & 1 & 8\\
    \midrule
    Average number or initiatives per response & 1.8 & 3.7 & 3.1 & 3.5 & 3.4 \\
  \end{tabularx}
  \label{tab:digital-initiatives}
 \end{table}

 Table \ref{tab:digital-initiatives} further reports average numbers of digital transformation initiatives by cluster. While leaders, experimenters and execution specialists hover around the average of 3.4 initiatives, the laggards report using less than two initiatives. With the average value of 3.7, the experimenters are slightly above average. The results can be interpreted in favour of the idea what level of organisational agility is connected with overall level of digitalisation. However, further research is needed to confirm this idea. Also, our data set does not include any hints on causation: does higher level of organisational agility lead to a higher level of digitalisation or vice versa.




\subsection{Agile at Scale}

Table \ref{tab:agile-at-scale} provides a summary statistics for the current use of agile scaling frameworks. Multiple answers were allowed.
76 per cent of respondents have confirmed deploying agile scaling frameworks. Scrum of Scrums and Lean Management are leading the list with 44 and 31 responses. Another 31 respondents have confirmed using an internally developed framework.
\begin{table}[hbt]
 \small
 \caption{Usage of agile scaling frameworks (breakdown responses by cluster)}
  \centering
  \begin{tabularx}{\textwidth}
  {Xlllll}
    Framework & Laggards & Experimenters & Execution specialists & Leaders & Total \\
    \midrule
    Scrum of Scrums & 7 & 19 & 4 & 14 & 44 \\
    Lean Management & 3 & 13 & 3 & 12 & 31 \\
    Internally created method & 6 & 9 & 4 & 12 & 31 \\
    SAFe & 4 & 8 & 2 & 9 & 23 \\
    LeSS & 2 & 2 & 2 & 7 & 13 \\
    Agile Portfolio Management & - & 3 & 3 & 7 & 13 \\
    Disciplined Agile Delivery & 1 & 1 & - & 3 & 5 \\
    Other & - & 1 & - & 2 & 3 \\
    We do not scale agile methods & 20 & - & 18 & 3 & 41 \\
     \midrule
  \end{tabularx}
  \label{tab:agile-at-scale}
 \end{table}

Table \ref{tab:number-scaling-frameworks} presents our results relating to the number of agile scaling frameworks in use at a time. The vast majority (62 responses) uses one or two agile scaling frameworks. Another 22 respondents revealed using three or more frameworks simultaneously. The sample average number of agile scaling frameworks equals 1.30 supporting the idea that, on average, the respondents tend to deploy one or two agile scaling framework at a time. However, the averages vary significantly by cluster. The Laggards and Execution Specialists have the lowest average of 0.66 and 0.60 respectively revealing that only every second respondent within those clusters uses agile scaling frameworks. The averages rise up to 2.20 for the Leaders and 1.87 for the Experimenters showing that those groups tend to simultaneously deploy two agile scaling frameworks.
\begin{table}[hbt]
 \small
 \caption{Number of agile scaling frameworks in current use}
  \centering
  \begin{tabular}{lcccccccc}
    & \multicolumn{6}{l}{Number of frameworks in use} \\
    \cmidrule{2-7}
    & 0 & 1 & 2 & 3 & 4 & 5 & Checksum & Average \\
    \midrule
    Laggards & 20 & 9 & 4 & 2 & - & - & 35 & 0.66 \\
    Execution specialists & 18 & 6 & 6 & - & - & - & 30 & 0.60 \\
    Experimenters & - & 12 & 11 & 6 & 1 & - & 30 & 1.87 \\
    Leaders & 3 & 8 & 6 & 9 & 1 & 3 & 30 & 2.20 \\
    \midrule
    Total & 41 & 35 & 27 & 17 & 2 & 3 & 125 & 1.30 \\
  \end{tabular}
  \label{tab:number-scaling-frameworks}
 \end{table}

\subsection{Business Agility}



Table \ref{tab:business-agility} highlights current perception of respondents with regard to the adoption of agile methods across individual organisational functions.
The respondents have been asked to select up to three most agile and least agile functions within their organisations. The number of responses is reported in the corresponding column of the table. All items are sorted in descending order by the value of the first column (most agile). The respondents seem to agree that information technology and product development behave in agile manner in their organisations. Indeed, 65 per cent of respondents confirm that IT is the most agile function, while only 15 per cent see IT as non-agile. Similarly, 54 per cent rate product development as agile, while only 6 per cent consider this function as non-agile.

Supporting functions such as human resources, corporate finance, general administration, as well as legal service, risk management and compliance are seen by the majority of the respondents as the least agile functions within their organisations. Only 4 per cent of respondents rate human resources, finance and administration as agile, while roughly 70 per cent confirm those functions being non-agile in their organisations.
Also, 24 per cent respondents rate the line management within their organisations as non-agile supporting the idea that there is still a need in facilitating agile leadership style and behaviours across organisations.

Respondents seem to be indifferent whether marketing and communications as well as customer service and support are currently more agile or non-agile. This observation may indicate that there is a lot of transition going on within those functions and there is no clear view of current results across organisations.
\begin{table}[hbt]
 \small
 \caption{Respondents' perception of the most and least agile organisational functions [count responses]}
  \centering
  \begin{tabular}{lll}
    Organisational function & Most agile & Least agile \\
    \midrule
    Information technology & 81 & 19 \\
    Product development & 68 & 7 \\
    Research & 48 & 10\\
    Production and operations & 35 & 26\\
    Marketing and communications & 21 & 23\\
    Customer service and support & 19 & 18\\
    Sales & 13 & 23\\
    HR, finance and administration & 5 & 87\\
    Legal, risk and compliance & 4 & 71\\
    Line management & 2 & 30\\
    \midrule
  \end{tabular}
  \label{tab:business-agility}
 \end{table}

\section{DISCUSSION}

We address the relationship between technical excellence, agile organisational design and agility. While the practice literature has encouraged managers to expect that organisational design changes enacted during agile transformation secure organisational agility, our findings suggest a more subtle relationship between technical excellence and agile organisational design. Four identified profiles (Leaders, Experimenters, Execution Specialists, and Laggards) suggest that organisational agility can be build upon technical excellence (Execution Specialists) and agile organisational design (Experimenters). However, combined in an intelligent way, both factors will create a consistent profile (Leaders).

When designing and implementing agile transformations, managing multiple organisational dimensions is critical for success. Understanding how the Leaders achieve organisational agility requires a nuanced appreciation of the link between agile organisational design, technical excellence, corporate culture and leadership styles. We argued that attempts to achieve greater organisational agility are associated with building more agile business processes rather than focusing on project work. Employees with agile mindset would voluntarily choose agile ways of working and shift project work into agile modes; however, the management has to establish the underlying framework of agile and lean processes to enable employees work effectively. 74 per cent of the Leaders agree that processes such as planning, budgeting and resource allocation are flexible enough to adjust to changing priorities, compared to 9 per cent of laggards. This finding runs counter to prescriptive literature and general managerial practice that advocate a greater reliance on agile project work to enable enact organisational agility. Though our data do not allow us to confirm underlying casual mechanisms, it is possible than agile business processes enable organisational to act in a more agile manner. Further research can more clearly delineate the reasons for this relationship. Technical excellence is mainly achieved through continuous delivery, deployment and integration, test automation and decoupled architectures. Focusing on product management rather than IT management allows the Leaders to consider the entire value chain by looking into activities how new products are created and existing products are modified.

When companies focus on designing agile transformations, managers actively seek to extend organisation by using agile scaling frameworks. In this context, scaling agile methods seem to be positively correlated with the level of organisational agility. Experimenters and Leaders reported to use around two frameworks simultaneously. This finding suggests that the advice in the practice literature on agile transformation design as a process of scaling agile practices is accurate. Companies that follow the traditional transformation approach, assuming that agile practices should be implemented on a team level first, may find themselves unable to scale agile practices up to the organisation-wide level. This result is more inline with the idea of designing agile transformation closely with an introduction of an agile scaling framework.

The respondents perceive that length of experience with agile practices is positively correlated with the level of organisational agility. Our findings suggest that a time frame of around two years is needed to achieve best possible impact from an agile transformation initiative. While managers seem to focus on increasing agility within IT, product development and research functions, a significant element of achieving agility stems from improving supporting functions. The respondents confirm supporting functions such as HR, finance, administration, legal, risk and compliance are currently run in a non-agile manner.



\section{ACKNOWLEDGEMENTS}

The survey presented in this paper has been funded and executed in close cooperation with kobaltblau Management Consultants. This collaboration offered a unique setting allowing to take advantage of the kobaltblau's expertise and broad industry network. The authors wish to thank kobaltblau for the invaluable contributions. The authors gratefully acknowledge extremely valuable discussions and support received from and Hans-Werner Feick, Thomas Heinevetter, Frederic Cuny, Christoph Hecker, Moritz Windelen, and Martin Tydecks. We also thank Alexander Rollinger and Natalie Roedenbeck for technical contributions and Jan Fikentscher for help with Tableau.


\section{ANNEX}

\subsection{Survey questions}
\label{subsec:survey-questions}

\begin{enumerate}
    \item How long has your organisation been using agile methods: fewer that 2, 2-4, 4-6, 6 years or more.
    \item What is the share of agile projects in your IT project portfolio: fewer that 20, 20-40, 40-60, 60 per cent or more.
    \item How effective would you say your organisation's agile transformation efforts have been to date: 0 = not effective at all, 10 = extremely effective (integer scale).
    \item Which dimensions of your organisation have been affected most by agile transformation? Please select up to five: organisational structure, processes, leadership, culture and values, goal setting, delivery and software development, software maintenance, product development, architecture, project management, demand management, governance, portfolio management.
    \item Please respond to each item in terms of how does it apply to your organisation (definitely true, probably true, neither true nor false, probably false, definitely false):
    \begin{enumerate}
        \item The management demonstrates leadership styles building upon employee empowerment, cross-functional collaboration and short feedback cycles.
        \item Agile values and principles are well known across the organisation.
        \item The organisation has established a positive failure attitude and embraces risk taking.
        \item Employees across the organisation have been equipped with substantial decision rights and exercise those rights.
        \item The management team has initiated organisational changes to further facilitate agile transformation.
        \item Project teams are staffed in a cross-functional manner and engage in cross-functional collaboration.
        \item The organisation has implemented new agile organisational models, e.g. value streams, virtual organisations.
    \end{enumerate}
    \item In your view, which areas of your organisation are most/least agile? Please select up to three for each column: IT, product development, research, production and operations, customer service and support, marketing and communications, sales, HR/finance/administration, legal/risk/compliance, line management.
    \item Please rate the extend to which you agree with each of the following statements (strongly disagree, somewhat disagree, neither disagree nor agree, somewhat agree, strongly agree):
    \begin{enumerate}
        \item In my organisation, tests are run in an automated manner and executed throughout the implementation phase.
        \item Continuous delivery, deployment and integration enable my organisation to deliver changes more frequently and reliably.
        \item My organisation has the ability to continuously incorporate customer feedback into the product development.
        \item Agile methods and tools are used for project-independent activities, e.g. maintenance, incident tracking, environment teams, value stream teams.
        \item When starting a new project, I can refer to decision criteria in my organisation on where and how to use agile methods.
        \item Planning, budgeting and resource allocation processes are flexible enough to fluidly adjust to changes in my organisation’s priorities.
        \item My organisation has established architecture principles supporting agile development through collaboration, emergent design, and design simplicity.
        \item Enterprise architecture is organisationally embedded into agile team structures.
    \end{enumerate}
    \item How would you rate your organisation's level of agile adoption relative to your competitors? (Much better, somewhat better, about the same, somewhat worse, much worse).
    \item Which of the following agile scaling frameworks do you use in your organisation? Please select all that apply: Scaled Agile Framework (SAFe), Scrum of Scrums, Lean Management, Agile Portfolio Management, Large-Scale Scrum (LeSS), Disciplined Agile Delivery (DAD), Recipes for Agile Governance in the Enterprise (RAGE), Nexus, internally created method, we don't scale agile methods.
    \item Which of the following can be found within your organisation? Please select all that apply: automated business processes, digital strategy, digital transformation projects and initiatives, digital products and services, digital business models, digital factory, chief digital officer, none from above.

\end{enumerate}

    \textbf{These last 4 questions are for classification purposes only. Please proceed.}

\begin{enumerate}
    \item How many individuals does your organisation employ (all locations): fewer than 500, 500-999, 1,000-4,999, 5,000-9,999, 10,000 or more.
    \item Which of the following best describes the industry sector in which you work: automotive, insurance, financial services (bank, asset management) excl. insurance, consumer goods, public sector, life sciences (pharmaceuticals, biotechnology), chemicals and materials, communications/media/entertainment, high tech, healthcare, energy and utilities, transport and logistics, other (free text).
    \item What is your main functional roles: business manager, IT manager, agile coach.
    \item Where is your organisation headquartered: Germany, France, Austria, Switzerland, UK, other (free text).
    \item Do you want us to share the survey results report with you? We will ask you to provide your contact information (name, position, company and email address). As soon as we have completed the survey, we will send you the download link to the survey results on your email address. By clicking on “I agree”, you give consent to the processing of your contact information. Your consent is entirely voluntary and can be withdrawn at any time, without giving of any reasons and with effect for the future. To withdraw your consent, please contact \texttt{datenschutz@kobaltblau.com}.
    \item Please provide your contact information for receiving the survey results report: name, position, company, email address (shown only if the respondent has given consent to data processing in the previous question).

\end{enumerate}

\printbibliography


\end{document}